\documentclass[11pt,twoside]{article}
\usepackage{asp2014}

\def \phcmsec{\hbox{ph cm$^{-2}$ s$^{-1}$}}

\aspSuppressVolSlug
\resetcounters
\begin{document}

\title{Preliminary Results of a New Deep Learning Method to Detect and Localize GRBs in the AGILE/GRID Sky Maps.}

\author{N.~Parmiggiani,$^1$ A.~Bulgarelli,$^1$ A.~Macaluso,$^2$ V.~Fioretti,$^1$ A.~Di~Piano,$^{1,3}$  L.~Baroncelli,$^1$ A.~Addis,$^1$ M.~Landoni,$^4$ C.~Pittori,$^{5,6}$ F.~Verrecchia,$^{5,6}$ F.~Lucarelli,$^{5,6}$ A.~Giuliani,$^7$  F.~Longo,$^8$ D.~Beneventano,$^3$ and M.~Tavani$^9$}

\affil{$^1$INAF OAS Bologna, Via P. Gobetti 93/3, 40129 Bologna, Italy. \email{nicolo.parmiggiani@inaf.it}}

\affil{$^2$German Research Center for Artificial Intelligence (DFKI), 66123 Saarbruecken, Germany.}

\affil{$^3$Universit\`{a} degli Studi di Modena e Reggio Emilia, DIEF, Via Pietro Vivarelli 10, 41125 Modena, Italy.}

\affil{$^4$INAF/OA Brera, via Brera 28, 20121 Milano, Italy.}

\affil{$^5$INAF/OAR Roma, Via Frascati 33, 00078 Monte Porzio Catone, Roma, Italy.}

\affil{$^6$ASI/SSDC Roma, Via del Politecnico snc, 00133 Roma, Italy.}

\affil{$^7$INAF/IASF Milano, Via Alfonso Corti 12, 20133 Milano, Italy.}

\affil{$^8$Dipartimento di Fisica, University of Trieste, and INFN, Trieste, via Valerio 2, 34127 Trieste, Italy}

\affil{$^9$INAF/IAPS Roma, Via del Fosso del Cavaliere 100, 00133 Roma, Italy.}


\paperauthor{Nicol\`{o}~Parmiggiani}{nicolo.parmiggiani@inaf.it}{0000-0002-4535-5329}{INAF}{OAS}{Bologna}{BO}{40129}{Italy}
\paperauthor{Andrea~Bulgarelli}{andrea.bulgarelli@inaf.it}{0000-0001-6347-0649}{INAF}{OAS}{Bologna}{BO}{40129}{Italy}
\paperauthor{Antonio~Macaluso}{antonio.macaluso.90@gmail.com }{0000-0002-1348-250X}{}{DFKI}{Saarbruecken}{66123}{}{Germany}
\paperauthor{Valentina~Fioretti}{valentina.fioretti@inaf.it}{0000-0002-6082-5384}{INAF}{OAS}{Bologna}{BO}{40129}{Italy}
\paperauthor{Ambra~Di~Piano}{ambra.dipiano@inaf.it}{0000-0002-9894-7491}{INAF}{OAS}{Bologna}{BO}{40129}{Italy}
\paperauthor{Leonardo~Baroncelli}{leonardo.baroncelli@inaf.it}{0000-0002-9215-4992}{INAF}{OAS}{Bologna}{BO}{40129}{Italy}
\paperauthor{Antonio~Addis}{antonio.addis@inaf.it}{0000-0002-0886-8045}{INAF}{OAS}{Bologna}{BO}{40129}{Italy}

\paperauthor{Marco~Landoni}{marco.landoni@inaf.it}{0000-0001-5570-5081}{INAF}{OA}{Brera}{MI}{20121}{Italy}

\paperauthor{Carlotta~Pittori}{carlotta.pittori@inaf.it}{0000-0001-6661-9779}{INAF}{OAR}{Monte Porzio Catone}{RO}{00078}{Italy}

\paperauthor{Francesco~Verrecchia}{francesco.verrecchia@inaf.it}{0000-0003-3455-5082}{INAF}{OAR}{Monte Porzio Catone}{RO}{00078}{Italy}

\paperauthor{Fabrizio~Lucarelli}{fabrizio.lucarelli@inaf.it}{0000-0002-6311-764X}{INAF}{OAR}{Monte Porzio Catone}{RO}{00078}{Italy}

\paperauthor{Andrea~Giuliani}{andrea.giuliani@inaf.it}{0000-0002-4315-1699}{INAF}{IASF}{Milano}{MI}{20133}{Italy}

\paperauthor{Francesco~Longo}{francesco.longo@inaf.it}{0000-0003-2501-2270}{University of Trieste}{Dipartimento di Fisica and INFN}{Trieste}{TS}{34127}{Italy}

\paperauthor{Domenico~Beneventano}{domenico.beneventano@unimore.it}{0000-0001-6616-1753}{Universit\`{a} degli Studi di Modena e Reggio Emilia}{DIEF}{Modena}{MO}{41125}{Italy}

\paperauthor{Marco~Tavani}{marco.tavani@inaf.it}{0000-0003-2893-1459}{INAF}{IAPS}{Monte Porzio Catone}{RO}{00078}{Italy}




\begin{abstract}

AGILE is an ASI space mission launched in 2007 to study X-ray and gamma-ray phenomena in the energy range from $\sim20$ keV to $\sim10$ GeV. The AGILE Team developed a real-time analysis pipeline for the fast detection of transient sources, and the follow-up of external science alerts received through networks such as the General Coordinates Network. We developed a new Deep Learning method for detecting and localizing Gamma-Ray Bursts (GRB) in the AGILE/GRID sky maps. We trained the model using sky maps with GRBs simulated in a radius of 20 degrees from the center of the map, which is larger than 99.5 \% of the error region present in the GRBWeb catalog. We also plan to apply this method to search for counterparts of gravitational wave events, which typically have a wider localization error region. The method comprises two Deep Learning models implemented with two Convolutional Neural Networks. The first model detects and filters sky maps containing a GRB, while the second model localizes its position. We trained and tested the models using simulated data. The detection model achieves an accuracy of 95.7 \%, and the localization model has a mean error lower than 0.8 degrees. We configured a Docker\footnote{https://www.docker.com/} container with all the required software for data simulation and deployed it using the Amazon Web Service\footnote{https://aws.amazon.com/} to calculate the p-value distribution under different conditions. With the p-value distribution, we can calculate the statistical significance of a detection.
  
\end{abstract}

\section{Introduction}

AGILE (Astrorivelatore Gamma ad Immagini LEggero - Light Imager for Gamma-Ray Astrophysics) is a space mission of the Italian Space Agency (ASI) devoted to high-energy astrophysics and launched on 23rd Apr 2007 \citep{2008NIMPA.588...52T, 2009A&A...502..995T}. The AGILE payload consists of the Silicon Tracker (ST), the SuperAGILE X-ray detector, the CsI(Tl) Mini-Calorimeter (MCAL), and an AntiCoincidence System (ACS). The combination of ST, MCAL, and ACS composes the Gamma-Ray Imaging Detector (GRID). This work aims to present a new detection method based on Deep Learning (DL, \citet{2015Natur.521..436L}) to identify and localize Gamma-Ray Bursts (GRBs) inside the AGILE/GRID intensity maps (counts maps divided by the exposure, in units of $\phcmsec sr^{-1}$). This new method will be implemented into the AGILE real-time analysis (RTA) pipelines \citep{2019ExA....48..199B, Parmiggiani:20214o} to detect transient phenomena during the follow-up of science alerts shared by other facilities through the General Coordinates Network (GCN, \url{https://gcn.nasa.gov/}). The AGILE Team has already used DL techniques to detect GRBs from AGILE/GRID intensity maps \citep{2021ApJ...914...67P}, however, in our first method, we could only detect GRBs within a radius of 1 degree from its center. In this new method, we improved the detection software by adding a new model that localizes the GRB in a radius of 20 degrees. This feature is essential to execute the follow-up of science alerts that have wide localization error regions. 

\section{Deep Learning Model}

We developed two DL models based on the Convolutional Neural Network (CNN) architecture \citep{Goodfellow-et-al-2016}, designed to extract features from images, using two open-source frameworks: Keras (\url{https://keras.io}) running on top of Tensorflow (\url{https://www.tensorflow.org}).
The first model classifies intensity maps to detect those which contain a GRB, and the second model localizes the GRB within a radius of 20 degrees. We decided to use a searching radius of 20 degrees after analyzing the GRBWeb catalog, which combines data from different sources and several detectors \citep{PhysRevD.102.103014},\url{https://user-web.icecube.wisc.edu/~grbweb_public/}, and finding that 99.5 \% of science alerts have an error radius smaller than 20 degrees. However, for events with wider uncertainty (e.g., gravitational wave alerts), a different analysis must be performed: a grid of blind searches must be executed to cover the full localization error region.

We analyzed several hundreds of different network configurations and selected the best performing ones. Both models have a first layer that receives input maps with a size of $100\times100$ pixels, then a first Convolution2D layer with filters of kernel size $12\times12$ pixels (to match the AGILE/GRID Point Spread Function or PSF) executes the convolution. All the convolutional layers use the Rectified Linear Unit (ReLU) activation function to improve computing efficiency. The next layer consists of a MaxPooling2D operation with a kernel size of $2\times2$ pixels to reduce the model size and speed up the training. From here, the two models implement a different number of additional Convolution2D layers and a MaxPooling2D layer after which a Dropout layer is added, with a probability of 20\% for the first model and 40 \% for the second. This layer is used as a regularization technique to prevent overfitting. The models implement then a Dense layer flattening the 2D data in a one-dimensional array, of 10 elements for the first model and ten thousand for the second one. Another Dropout layer follows. 

The final layer differs from the two models, since they have different goals. In fact, the first model (binary classifier) has a final layer composed of two neurons using a Softmax activation function that provides the probabilities of two classes: background and GRB intensity map. While the second model, which executes regression for the coordinates of the GRBs, implements a final layer with two neurons using a Sigmoid activation function. 

Both DL models are trained with a supervised technique. For this reason, we executed Monte Carlo simulations and generated two datasets to train and test the two models,  each with 40000 AGILE-GRID intensity sky maps. The simulation procedure labels whether or not a simulated GRB is present and its position. We apply a Gaussian smoothing of 6 degrees in radius on the simulated maps, assuming twice the value of the AGILE-GRID PSF for the energy range considered. The source models and the background conditions are the same as \citet{2021ApJ...914...67P}. The GRBs are simulated in a random position within 20 degrees from the center of the map to simulate external science alerts. The balanced datasets contain half of the maps containing a GRB and the other half background-only. No additional sources are simulated.

\section{Model Training and Preliminary Results}

\begin{figure*}[!htb]
	\centering
	  \includegraphics[width=0.45\textwidth]{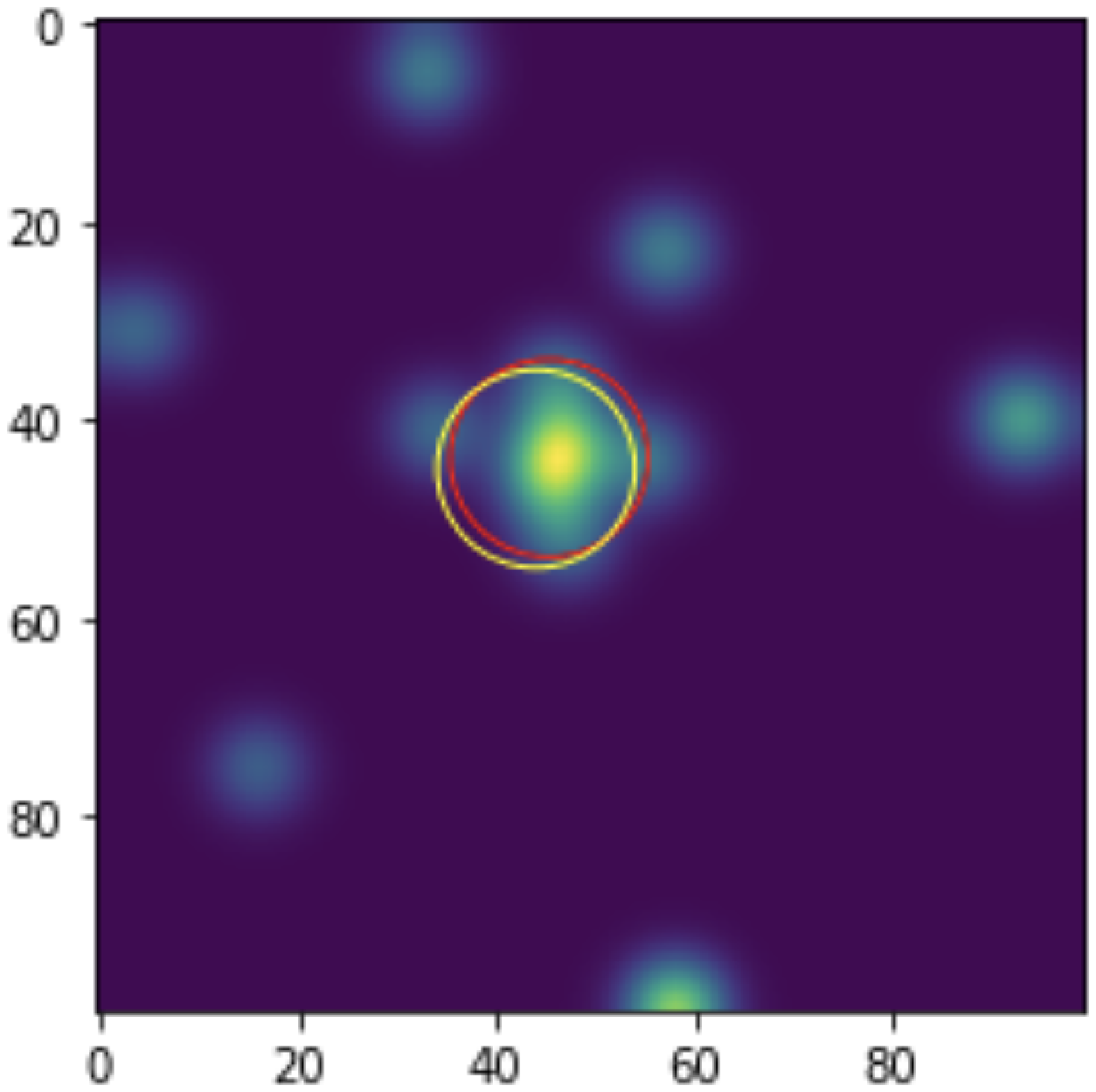}
	  \includegraphics[width=0.45\textwidth]{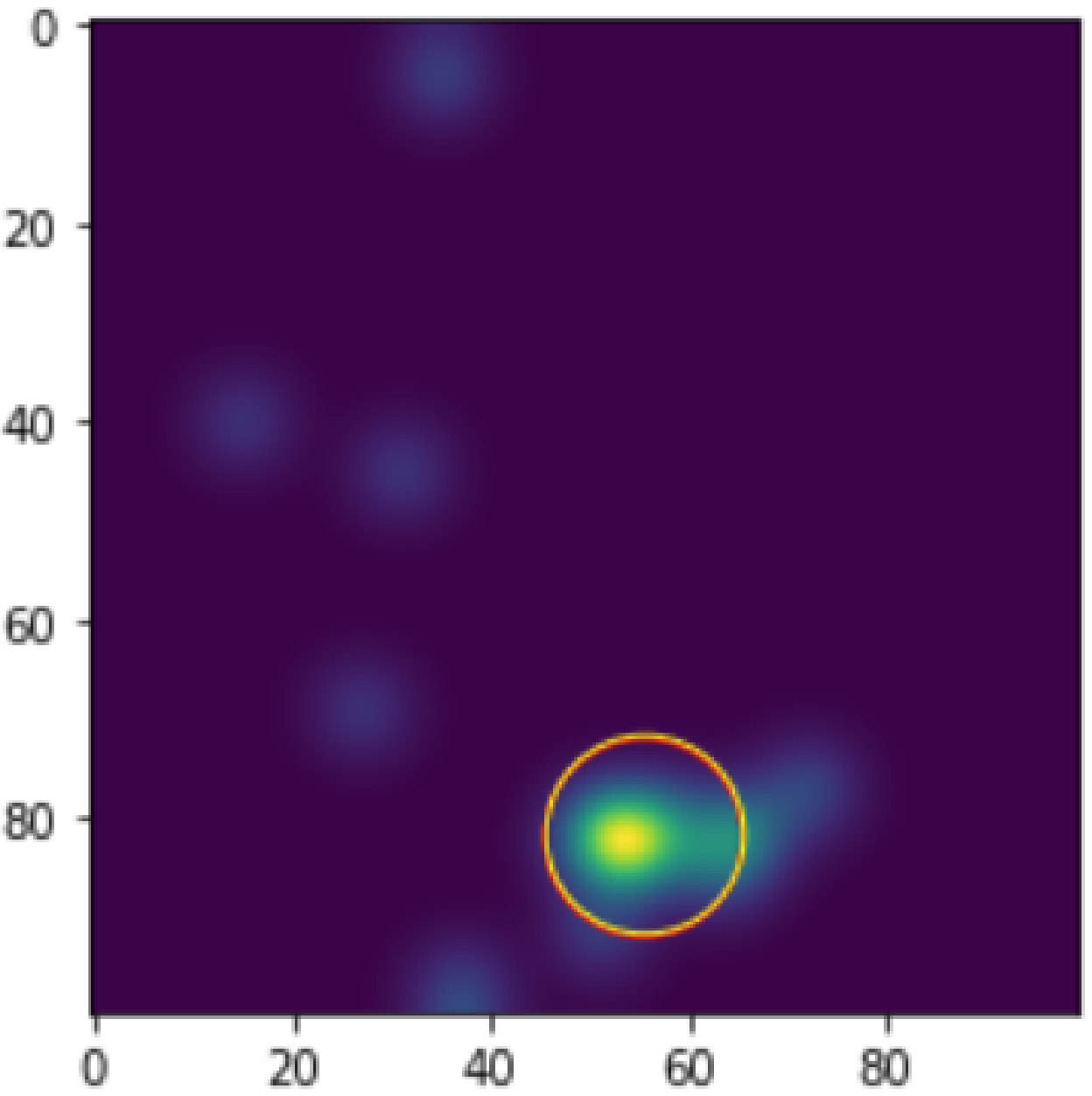}
	\caption{Intensity maps with a simulated GRB (yellow circle). The predicted GRB position is marked with a red circle. The axes represent the pixel of the maps.}
	\label{fig:loc_results}
\end{figure*}

We performed the training of the first model with a batch size of 200 maps for 10 epochs. The reconstruction error is computed with a sparse categorical crossentropy loss function. The accuracy reached after ten epochs is equal to 95.7 \% on the test dataset. The training of the second model is performed with a batch size of 200 maps for 26 epochs. The reconstruction error is computed with a mean absolute error. The mean localization accuracy is lower than 0.8 degrees on the test dataset. Figure \ref{fig:loc_results} shows examples of sky maps containing GRBs; the simulated position is marked with a yellow circle whilst the predicted position with a red circle. Both models use the ADAM optimization algorithm \citep{2014arXiv1412.6980K} configured with a learning rate of 0.001.

\section{Conclusions and Future Works}

We developed two DL models based on the CNN architecture to detect and localize GRBs in the AGILE/GRID intensity maps, within a radius of 20 degrees from the center. The aim is to implement these models in the AGILE RTA pipeline for the follow-up of external science alerts, regardless of their localization error region. A first model classifies maps containing a GRB, while the second localizes its position. We trained the two models with a balanced dataset comprising maps with background-only and maps with a simulated GRB randomly positioned within 20 degrees from the center. The first model achieves 95.7 \% of classification accuracy, and the second model achieves a mean localization error lower than 0.8 degrees. We prepared a Docker container installing the required software for the AGILE/GRID map simulation. By computing the p-value distribution, we could calculate the statistical significance of a GRB detection. We deployed the container in the Amazon Web Service infrastructure to simulate millions of maps, exploiting its cloud computing capabilities.

The method presented in this work can be used to develop DL models for other gamma-ray space missions in a similar energy range. 

\acknowledgements The AGILE Mission is funded by the Italian Space Agency (ASI) with scientific and programmatic participation by the Italian National Institute for Astrophysics (INAF) and the Italian National Institute for Nuclear Physics (INFN). The investigation is supported by the ASI grant  I/028/12/6. We thank the ASI management for unfailing support during AGILE operations. We acknowledge the effort of ASI and industry personnel at the ASI ground station in Malindi (Kenya), at the Telespazio Mission Control Center at Fucino, and the data processing done at the ASI/SSDC in Rome: the success of AGILE scientific operations depends on the effectiveness of the data flow from Kenya to SSDC and the data analysis and software management.


\bibliography{P46}


\end{document}